\documentclass{amsart}

\usepackage{epsfig,amsfonts,amssymb,amsmath,amsthm,eucal,pinlabel}
\usepackage[all]{xy}

\newtheorem{thm}{Theorem}[section]

\newtheorem{prop}[thm]{Proposition}
\newtheorem{cor}[thm]{Corollary}
\theoremstyle{remark}\newtheorem*{rem}{Remark}
\theoremstyle{remark}

\newenvironment{romanlist}
        {\begin{enumerate}
        }
        {\end{enumerate}}

\newcounter{ticklistc}
\newenvironment{ticklist}
    	{\setcounter{ticklistc}{0}
	 \begin{list}{--}
	{\usecounter{ticklistc}}}{\end{list}}

\newcommand{\Z}{\mathbb Z}
\newcommand{\R}{\mathbb R}
\newcommand{\C}{\mathbb C}

\newcommand{\K}{\mathcal K}

\newcommand{\E}{\mathcal E}
\newcommand{\SI}{\Sigma}
\newcommand{\G}{\Gamma}
\newcommand{\GSS}{\Gamma\subset\Sigma}

\newcommand{\e}{\varepsilon}

\newcommand{\Pf}{\mbox{Pf}}

\newcommand{\D}{\mathcal D}

\newcommand{\A}{\mbox{Arf}}

\begin{document}

\title{Dimers on surface graphs and spin structures. II}

\author{David Cimasoni}
\address{Department of Mathematics, UC Berkeley, 970 Evans Hall, Berkeley, CA 94720, USA}
\email{cimasoni@math.berkeley.edu}

\author{Nicolai Reshetikhin}
\email{reshetik@math.berkeley.edu}


\date{\today}

\begin{abstract}
In a previous paper \cite{C-R}, we showed how certain orientations of the edges of a graph $\G$ embedded in a closed oriented surface $\SI$
can be understood as discrete spin structures on $\SI$. We then used this correspondence to give a geometric proof of
the Pfaffian formula for the partition function of the dimer model on $\G$. In the present article, we generalize these results to the case
of compact oriented surfaces with boundary. We also show how the operations of cutting and gluing act on discrete spin structures and how they
change the partition function. These operations allow to reformulate the dimer model as a quantum field theory on surface graphs.
\end{abstract}

\maketitle

\tableofcontents

\section*{Introduction}

A dimer configuration on a graph $\G$ is a choice of a family of edges of $\G$, called dimers, such that each vertex of $\G$ is adjacent
to exactly one dimer. Assigning weights to the edges of $\G$ allows to define a probability measure on the set of
dimer configurations. The study of this measure is called the dimer model on $\G$.
Dimer models on graphs have a long history in statistical mechanics \cite{Kast1,McCoyW}, but also show
interesting aspects involving combinatorics, probability theory \cite{Kup,CKP},
real algebraic geometry \cite{KenOkS,KenOk}, etc...

A remarkable fact about dimer models was discovered by P.W. Kasteleyn in the 60's: the partition function of the dimer model
can be written as a linear combination of $2^{2g}$ Pfaffians of $N\times N$ matrices, where $N$ is the number of
vertices in the graph and $g$ the genus of a closed oriented surface $\SI$ where the graph can be embedded.
The matrices are signed-adjacency matrices, the sign being determined by an orientation of the edges of $\G$ called a
Kasteleyn orientation. If the graph is embedded in a surface of genus $g$, there are exactly $2^{2g}$ equivalence classes of
Kasteleyn orientations, defining the $2^{2g}$ matrices. This Pfaffian formula for the partition function was proved by Kasteleyn in \cite{Kast1}
for the cases $g=0,1$, and only stated for the general case \cite{Kast2}. A combinatorial proof of this fact and the
exact description of coefficients for all oriented surfaces first appeared much later \cite{L,T}.

The number of equivalence classes of Kasteleyn orientations on a graph $\G$ embedded in $\Sigma$ is also equal to the number of
equivalence classes of spin structures on $\SI$. An explicit construction relating a spin structure on a surface
with a Kasteleyn orientation on a graph with dimer configuration was suggested in
\cite{Kup}. In \cite{C-R}, we investigated further the relation between
Kasteleyn orientations and spin structures. This allows to understand Kasteleyn orientations on a graph embedded in $\SI$
as discrete spin structures on $\SI$. We also used this relation
to give a geometric proof of the Pfaffian formula for closed surfaces. Our final formula can be expressed as follows:
given a graph $\G$ embedded in a closed oriented surface $\SI$ of genus $g$, the partition function
of the dimer model on $\G$ is given by
\[
Z(\G)=\frac{1}{2^g}\sum_{\xi\in\mathcal{S}(\SI)}\A(\xi)\Pf(A^\xi(\G)),
\]
where $\mathcal{S}(\SI)$ denotes the set of equivalence classes of spin structures on $\SI$, $\A(\xi)=\pm 1$ is the
Arf invariant of the spin structure $\xi$, and $A^\xi(\G)$ is the matrix given by the Kasteleyn orientation corresponding to $\xi$.

\medskip

The first part of the present paper is devoted to the extension of the results obtained in
\cite{C-R} to dimer models on graphs embedded in surfaces with
boundary (Sections \ref{section:dimer} and \ref{section:Kasteleyn}).
We then show how the operations of
cutting and gluing act on discrete spin structures and how they
change the partition function (Section \ref{section:glue}).
These operations define the structure
of a functorial quantum field
theory in the spirit of \cite{At,Segal}, as detailed in Section \ref{section:QFT}.
We then give two equivalent
reformulations of the dimer quantum field theory: the ``Fermionic" version, which describes the
partition function of the dimer model as a Grassman integral,
and the ``Bosonic" version, the equivalent description of
dimer models on bipartite surface graphs in terms of height
functions. This special case of bipartite graphs is the subject of
Section \ref{section:bipartite}.

Throughout this paper, $\Sigma$ is a compact surface, possibly disconnected and possibly with boundary,
endowed with the counter-clockwise orientation. All results can be
extended to the case of non-orientable surfaces, which will be
done in a separate publication. We refer to \cite{T} for a combinatorial treatment of
dimer models on non-orientable surface graphs.

\subsection*{Acknowledgements}

We are grateful to J. Andersen, M. Baillif, P. Teichner and A. Vershik for inspiring discussions.
We also thankfully acknowledge the hospitality of the Department of Mathematics at the University of Aarhus.
The work of D.C. was supported by the Swiss National Science Foundation.
This work of N.R. was partially supported by the NSF grant DMS--0307599, by the CRDF grant RUM1--2622,
by the Humboldt foundation and by the Niels Bohr research grant.

\section{The dimer model on graphs with boundary}
\label{section:dimer}

\subsection{Dimers on graphs with boundary}
\label{sub:dimer-graph}

In this paper, a {\em graph with boundary\/} is a finite graph $\G$ together with
a set $\partial\G$ of one valent vertices called {\em boundary vertices\/}.
A {\em dimer configuration\/} $D$ on a graph with boundary $(\G,\partial\G)$
is a choice of edges of $\G$, called {\em dimers\/}, such
that each vertex that is not a boundary vertex is adjacent to exactly one dimer. Note that
some of the boundary vertices may be adjacent to a dimer of $D$,
and some may not. We shall denote by $\partial D$ this partition
of boundary vertices into matched and non-matched. Such a partition
will be called a {\em boundary condition\/} for dimer
configurations on $\G$.

A {\em weight system\/} on $\G$ is a positive real
valued function $w$ on the set of edges of $\G$. It defines edge
weights on the set $\D(\G,\partial\G)$ of dimer configurations on $(\G,\partial\G)$ by
\[
w(D)=\prod_{e\in D} w(e),
\]
where the product is taken over all edges occupied by dimers of $D$.

Fix a boundary condition $\partial D_0$. Then, the Gibbs measure
for the dimer model on $(\G,\partial\G)$ with weight system $w$
and boundary condition $\partial D_0$ is given by
\[
\mbox{Prob}(D\,|\,\partial D_0)=\frac{w(D)}{Z(\G; w\,|\,\partial D_0)},
\]
where
\[
Z(\G; w\,|\,\partial D_0)=\sum_{D:\partial D=\partial D_0} w(D),
\]
the sum being on all $D\in \D(\G,\partial\G)$ such that $\partial D=\partial D_0$.

Let $V(\G)$ denote the set of vertices of $\G$. The group
\[
{\mathcal G}(\G)=\{s\colon V(\G)\to\R_{>0}\}
\]
acts on the set of weight systems on $\G$ as follows: $(sw)(e)=s(e_+)w(e)s(e_-)$, where $e_+$ and $e_-$ are the two vertices adjacent to
the edge $e$. Note that $(sw)(D)=\prod_{v}s(v)w(D)$ and
$Z(\G;sw\,|\,\partial D_0)=\prod_{v}s(v)Z(\G;w\,|\,\partial D_0)$, both products being on the set of vertices of $\G$ matched
by $D_0$. Therefore, the Gibbs measure is invariant under the action of the group ${\mathcal G}(\G)$.

Note that the dimer model on $(\G,\partial\G)$ with boundary
condition $\partial D_0$ is equivalent to the dimer model on the
graph obtained from $\G$ by removing all edges adjacent to
non-matched boundary vertices.

\medskip

Given two dimer configurations $D$ and $D'$ on a graph with boundary $(\G,\partial\G)$, let us define the {\em $(D,D')$-composition cycles\/}
as the connected components of the symmetric difference $C(D,D')=(D\cup D')\backslash(D\cap D')$.
If $\partial D=\partial D'$, then $C(D,D')$ is a 1-cycle in $\G$ with $\Z_2$-coefficients. In general, it is only a 1-cycle $(rel\;\partial\G)$.

\subsection{Dimers on surface graphs with boundary}
\label{sub:dimer-surface}

Let $\Sigma$ be an oriented compact surface, not necessarily
connected, with boundary $\partial\Sigma$. A {\em surface graph
with boundary} $\G\subset \SI$ is a graph with boundary $(\G,\partial \G)$ embedded in $\SI$, so that
$\G\cap\partial\SI=\partial\G$ and the complement of $\G\setminus\partial\G$ in $\SI\setminus\partial\SI$ consists of open 2-cells.
These conditions imply that the graph $\overline\G:=\G\cup\partial\SI$ is the $1$-skeleton of a cellular
decomposition of $\SI$.

Note that any graph with boundary can be realized as a surface graph with boundary. One way is to embed the graph in a closed surface of
minimal genus, and then to remove one small open disc from this surface near each boundary vertex of the graph.

A {\em dimer configuration} on a surface graph with boundary $\G\subset \SI$ is simply a dimer
configuration on the underlying graph with boundary $(\G, \partial\G)$.
Given two dimer configurations $D$ and $D'$ on a surface graph $\GSS$,
let $\Delta(D,D')$ denote the homology class of $C(D,D')$ in $H_1(\SI,\partial\SI;\Z_2)$. We shall say that two dimer configurations
$D$ and $D'$ are {\em equivalent\/} if $\Delta(D,D')=0\in H_1(\SI,\partial\SI;\Z_2)$.
Note that given any three dimer configurations $D,D'$, and $D''$
on $\G\subset \SI$, we have the identity
\begin{equation}\label{add-compc}
\Delta(D,D')+\Delta(D',D'')=\Delta(D, D'')
\end{equation}
in $H_1(\SI, \partial \SI; \Z_2)$.

Fix a homology class $\beta\in H_1(\SI, \partial \SI; \Z_2)$, a
dimer configuration $D_1\in \D(\G, \partial \G)$ and a boundary
condition $\partial D_0$. The associated partial partition
function is defined by
\[
Z_{\beta, D_1}(\G;w\,|\,\partial D_0)=\sum_{\genfrac{}{}{0pt}{}{D:\partial D=\partial D_0}{\Delta(D,D_1)=\beta}}w(D),
\]
where the sum is taken over all $D\in \D(\G, \partial \G)$ such that
$\partial D=\partial D_0$ and $\Delta(D,D_1)=\beta$.

The equality  (\ref{add-compc}) implies that
\[
Z_{\beta, D_1}(\G; w\,|\,\partial D_0)=Z_{\beta+\Delta(D_0, D_1), D_0}(\G; w\,|\,\partial D_0).
\]
Furthermore, the relative homology class
$\beta'=\beta+\Delta(D_0,D_1)$ lies in the image of the canonical
homomorphism $j: H_1(\SI, \Z_2)\to H_1(\SI, \partial \SI; \Z_2)$.
Hence,
\[
Z_{\beta', D_0}(\G; w\,|\,\partial D_0)=\sum_{\alpha: j(\alpha)=\beta'}Z_\alpha(\G, w\,|\,\partial D_0),
\]
where the sum is taken over all $\alpha\in H_1(\SI, \Z_2)$ such
that $j(\alpha)=\beta'$, and
\[
Z_\alpha(\G;w\,|\,\partial D_0)=\sum_{\genfrac{}{}{0pt}{}{D:\partial D=\partial D_0}{\Delta(D,D_0)=\alpha}}w(D).
\]
Therefore the computation of the partition function $Z_{\beta, D_1}(\G; w\,|\,\partial D_0)$
boils down to the computation of  $Z_\alpha(\G;w\,|\,\partial D_0)$ with
$\alpha\in H_1(\SI;\Z_2)$. We shall give a Pfaffian formula for this latter partition function
in the next section (see Theorem \ref{thm:Pfaffian}).

\section{Kasteleyn orientations on surface graphs with boundary}
\label{section:Kasteleyn}

\subsection{Kasteleyn orientations}
\label{sub:Kasteleyn}

Let $K$ be an orientation of the edges of a graph $\G$, and let $C$ be an oriented closed curve in $\G$. We shall denote by
$n^K(C)$ the number of times that, traveling once along $C$ following its orientation, one runs along an edge in the direction
opposite to the one given by $K$.

A {\em Kasteleyn orientation\/} on a surface graph with boundary $\GSS$ is an orientation $K$ of the edges of
$\overline\G=\Gamma\cup\partial\SI$ which satisfies the following condition: for each face $f$ of $\SI$, $n^K(\partial f)$ is odd.
Here $\partial f$ is oriented as the boundary of $f$, which inherits the orientation of $\SI$.

Using the proof of \cite[Theorem 3.1]{C-R}, one easily
checks that if $\partial\SI$ is non-empty, then there always
exists a Kasteleyn orientation on $\GSS$. More precisely, we have
the following:

\begin{prop}
Let $\Gamma\subset\Sigma$ be a connected surface graph, possibly
with boundary, and let $C_1,\dots,C_\mu$ be the boundary
components of $\Sigma$ with the induced orientation. Finally, let
$n_1,\dots,n_\mu$ be $0$'s and $1$'s. Then, there exists a Kasteleyn
orientation on $\Gamma\subset\Sigma$ such that $1+n^K(-C_i)\equiv n_i\pmod{2}$
for all $i$ if and only if
$$
n_1+\cdots+n_\mu\equiv V\pmod{2},
$$
where $V$ is the number of vertices of $\Gamma$.
\end{prop}
\begin{proof}
First, let us assume that there is a Kasteleyn orientation $K$ on $\Gamma\subset\Sigma$ such
that $1+n^K(-C_i)\equiv n_i$ for all $i$. Let $\Sigma'$ be the closed
surface obtained from $\Sigma$ by pasting a 2-disc $D_i$ along each boundary component $C_i$.
Let $\Gamma'\subset\Sigma'$ be the surface graph obtained
from $\overline\Gamma$ as follows: for each $i$ such that $n_i=1$, add one vertex in the interior of $D_i$
and one edge (arbitrarily oriented) between this
vertex and a vertex of $C_i$. The result is a Kasteleyn orientation on $\Gamma'\subset\Sigma'$,
with $\Sigma'$ closed. By \cite[Theorem 3.1]{C-R},
the number $V'$ of vertices of $\Gamma'$ is even. Hence,
$$
0\equiv V'\equiv V+n_1+\cdots+n_\mu\pmod{2}.
$$
Conversely, assume $\Gamma\subset\Sigma$ is a surface graph with $n_1+\cdots+n_\mu\equiv V\pmod{2}$.
Paste $2$-discs along the boundary components of
$\Sigma$ as before. This gives a surface graph $\Gamma'\subset\Sigma'$ with $\Sigma'$ closed
and $V'$ even. By \cite[Theorem 3.1]{C-R}, there exists
a Kasteleyn orientation $K'$ on $\Gamma'\subset\Sigma'$. It restricts to a Kasteleyn orientation
$K$ on $\Gamma\subset\Sigma$ with
$1+n^K(-C_i)\equiv n_i$ for all $i$.
\end{proof}

Recall that two Kasteleyn orientations are called {\em equivalent\/} if one can be obtained from
the other by a sequence of moves reversing orientations of all edges adjacent to a vertex.
The proof of \cite[Theorem 3.2]{C-R} goes through verbatim: if non-empty, the set of equivalence classes
of Kasteleyn orientations on $\GSS$ is an affine $H^1(\Sigma;\Z_2)$-space.
In particular, there are exactly $2^{b_1(\SI)}$ equivalence classes of
Kasteleyn orientations on $\GSS$.

\subsection{Discrete spin structures}
\label{sub:discrete}

As in the closed case, any dimer configuration $D$ on a graph $\G$ allows to identify
equivalence classes of Kasteleyn orientations on $\GSS$ with spin structures on $\SI$. Indeed,
\cite[Theorem 4.1]{C-R} generalizes as follows.

Given an oriented simple closed curve $C$ in $\overline\G$, let $\ell_D(C)$ denote the number of vertices $v$ in $C$
whose adjacent dimer of $D$ sticks out to the left of $C$ in $\Sigma$. Also, let $V_{\partial D}(C)$ be the number of
boundary vertices $v$ in $C$ not matched by $D$, and such that the interior of $\Sigma$ lies to the right of $C$ at $v$.

\begin{thm}\label{thm:spinb}
Fix a dimer configuration $D$ on a surface graph with boundary $\GSS$.
Given a class $\alpha\in H_1(\SI;\Z_2)$, represent it by oriented simple closed curves $C_1,\dots,C_m$ in $\overline\G$.
If $K$ is a Kasteleyn orientation on
$\GSS$, then the function $q^K_D\colon H_1(\Sigma;\Z_2)\to\Z_2$ given by
\[
q^K_D(\alpha)=\sum_{i<j}C_i\cdot C_j+\sum_{i=1}^m(1+n^K(C_i)+\ell_D(C_i)+V_{\partial D}(C_i))\pmod{2}
\]
is a well-defined quadratic form on $H_1(\Sigma;\Z_2)$.
\end{thm}
\begin{proof}
Fix a dimer configuration $D$ on $(\G,\partial\G)$ and a Kasteleyn orientation
$K$ on $\GSS$. Let $\SI'$ be the surface (homeomorphic to $\SI$) obtained
from $\SI$ by adding a small closed collar to its boundary. For every vertex $v$ of $\partial\G$
that is not matched by a dimer of $D$, add a vertex $v'$
near $v$ in the interior of the collar and an edge between $v$ and $v'$.
Let us denote by $\G'$ the resulting graph in $\SI'$. Putting a dimer on each of
these additional edges, and orienting them arbitrarily, we obtain a perfect matching $D'$
and an orientation $K'$ on $\G'$. Although $\G'\subset\SI'$ is
not strictly speaking a surface graph, all the methods of \cite[Section 4]{C-R} apply. Indeed,
Kuperberg's vector field defined near $\G'$ clearly extends continuously to the collar. As in the closed case,
it also extends to the faces with even
index singularities. Using the perfect matching $D'$ on $\G'$, we obtain a vector
field $f(K',D')$ with even index singularities, which determines a spin structure
$\xi_{f(K',D')}$ on $\SI'$. Johnson's theorem \cite{Jo} holds for surfaces with boundary, so this spin
structure defines a quadratic form $q$ on $H_1(\SI';\Z_2)= H_1(\SI;\Z_2)$. If $C$ is a simple close curve
in $\G'\subset\SI'$, then $q([C])+1=n^{K'}(C)+\ell_{D'}(C)$ as in the closed case. The proof is completed using the
equalities $n^{K'}(C)=n^{K}(C)$ and $\ell_{D'}(C)=\ell_{D}(C)+V_{\partial D}(C)$.
\end{proof}

Since Johnson's theorem holds true for surfaces with boundary and \cite[Proposition 4.2]{C-R}
easily extends,
we have the following corollary.

\begin{cor}\label{cor:spin}
Let $\GSS$ be a surface graph, non-necessarily connected, and possibly with boundary.
Any dimer configuration $D$ on $\GSS$ induces an isomorphism of affine $H^1(\Sigma;\Z_2)$-spaces
\[
\psi_D\colon\K(\GSS)\longrightarrow {\mathcal{S}}(\SI)
\]
from the set of equivalence classes of Kasteleyn orientations on $\GSS$ onto the set of spin structures on $\SI$.
Furthermore, $\psi_D-\psi_{D'}$ is equal to the Poincar\'e dual of $\Delta(D,D')$. In particular, $\psi_D=\psi_{D'}$
if and only if $D$ and $D'$ are equivalent dimer configurations.\qed
\end{cor}

\subsection{The Pfaffian formula for the partition function}
\label{sub:Pfaffian}

Let $\Gamma$ be a graph, not necessarily connected, and possibly with boundary, endowed with a weight system $w$.
Realize $\G$ as a surface graph $\GSS$, and fix a Kasteleyn orientation $K$ on it. The {\em Kasteleyn coefficient\/} associated
to an ordered pair $(v,v')$ of distinct vertices of $\G$ is the number
\[
a^K_{vv'}=\sum_{e}\e_{vv'}^K(e)w(e),
\]
where the sum is on all edges $e$ in $\overline\Gamma$ between the vertices $v$ and $v'$, and
\[
\e^K_{vv'}(e)=
\begin{cases}
\phantom{-}1 & \text{if $e$ is oriented by $K$ from $v$ to $v'$;} \\
-1 & \text{otherwise.}
\end{cases}
\]
One also sets $a^K_{vv}=0$.
Let us fix a boundary condition $\partial D_0$ and enumerate the matched vertices of $\G$ by $1,2,\dots, 2n$.
Then, the corresponding coefficients form a $2n\times 2n$ skew-symmetric matrix $A^K(\overline\Gamma;w\,|\,\partial D_0)=A^K$ called
the {\em Kasteleyn matrix\/}.

Let $D$ be a dimer configuration on $(\Gamma,\partial\G)$ with $\partial D=\partial D_0$,
given by edges $e_1,\dots,e_n$ matching vertices $i_\ell$ and $j_\ell$ for $\ell=1,\dots,n$.
Let $\sigma$ be the permutation $(1,\dots, 2n)\mapsto (i_1,j_1,\dots,i_n,j_n)$, and set
\[
\e^K(D)=(-1)^\sigma\prod_{\ell=1}^n\e^K_{i_\ell j_\ell}(e_\ell),
\]
where $(-1)^\sigma$ denotes the sign of $\sigma$. Note that $\e^K(D)$ does not depend on the choice of $\sigma$,
but only on the dimer configuration $D$.

Finally, recall that the {\em Arf invariant\/} of a (possibly degenerate) quadratic form $q$ on $H:=H_1(\Sigma;\Z_2)$ is defined by
\[
\A(q)=\frac{1}{|H|}\sum_{\alpha\in H}(-1)^{q(\alpha)}.
\]
If there is a component $\gamma$ of $\partial\SI$ such that $q(\gamma)\neq 0$, then one easily checks that $\A(q)=0$.
On the other hand, if $q(\gamma)=0$ for all boundary components $\gamma$ of $\SI$, then 
$\A(q)$ takes the values $+1$ or $-1$.

\begin{thm}\label{thm:Pfaffian}
Let $\Gamma\subset\Sigma$ be a surface graph, not necessarily connected, and possibly with boundary.
Let $b_1(\Sigma)$ denote the dimension of $H_1(\Sigma;\Z_2)$, and let $g$ denote the genus of $\Sigma$.
Then,
\[
Z_\alpha(\Gamma;w\,|\,\partial D_0)=\frac{1}{2^{b_1(\Sigma)}}\sum_{[K]}(-1)^{q^K_{D_0}(\alpha)}\e^K(D_0)\Pf(A^K)
\]
for any $\alpha\in H_1(\Sigma;\Z_2)$, and
\[
Z(\Gamma;w\,|\,\partial D_0)=\frac{1}{2^g}\sum_{[K]}\A(q_{D_0}^K)\e^K(D_0)\Pf(A^K),
\]
where both sums are over the $2^{b_1(\Sigma)}$ equivalence classes of Kasteleyn orientations on $\GSS$.
Furthermore, $\A(q_{D_0}^K)\e^K(D_0)$ does not depend on $D_0$.
\end{thm}
\begin{proof}
First note that if the theorem holds for two surface graphs, then it holds for their disjoint union. Therefore, it may be assumed that $\Sigma$
is connected. The first formula follows from Theorem \ref{thm:spinb}: the proof of Theorem 4 and the first half of the proof of Theorem 5 of \cite{C-R}
generalize verbatim to the case with (possible) boundary.
The second formula can be obtained from the first one by summing over all $\alpha\in H_1(\Sigma;\Z_2)$.
However, this requires some cumbersome computations, so let us give another proof of this equality.
As mentioned in Section \ref{section:dimer}, the dimer model on $(\G,\partial\G)$ with boundary condition $\partial D_0$ is equivalent to
the dimer model on the graph $\G'=\G'(\partial D_0)$ obtained from $\G$ by removing all edges adjacent to non-matched boundary vertices.
Let $w'$ denote the restriction of $w$ to $\G'$. If $\GSS$ is a surface graph with boundary, then $\G'\subset\SI'$ is a surface graph,
where $\SI'$ is the closed oriented surface obtained from $\SI$ by gluing discs along all boundary components. By \cite[Theorem 5.3]{C-R},
\[
Z(\G;w\,|\,\partial D_0)=Z(\G';w')=\frac{1}{2^g}\sum_{[K']}\A(q^{K'}_{D_0})\e^{K'}(D_0)\Pf(A^{K'}(\G';w')),
\]
the sum being on all equivalence classes of Kasteleyn orientations on $\G'\subset\SI'$. Such a Kasteleyn orientation $K'$ extends uniquely to
a Kasteleyn orientation $K$ on $\GSS$ such that $q^K_{D_0}(\gamma)=0$ for all boundary component $\gamma$ of $\SI$. Furthermore,
$\e^{K'}(D_0)=\e^K(D_0)$ and $A^{K'}(\G';w')=A^K(\overline\G;w\,|\,\partial D_0)$. Since $\A(q^K_{D_0})=0$ for all other Kasteleyn
orientations, the theorem follows.
\end{proof}

\section{Cutting and gluing}
\label{section:glue}

\subsection{Cutting and gluing graphs with boundary}
\label{sub:glue-graph}

Let $(\G,\partial\G)$ be a graph with boundary, and let us fix an edge $e$ of $\G$. Let $(\G_{\{e\}},\partial\G_{\{e\}})$
denote the graph with boundary obtained from $(\G,\partial\G)$ as follows: cut the edge $e$ in two, and set
$\partial\G_{\{e\}}=\partial\G\cup\{v',v''\}$, where $v'$ and $v''$ are the new one valent vertices.
Iterating this procedure for some set of edges $\E$ leads to a graph with boundary $(\G_\E,\partial\G_\E)$, which is said to be obtained by
{\em cutting $(\G,\partial\G)$ along $\E$}.

Note that a dimer configuration $D\in \D(\G,\partial\G)$ induces
an obvious dimer configuration $D_\E\in \D(\G_\E,\partial\G_\E)$: cut in two the dimers of $D$ that belong to $\E$.

A weight system $w$ on $\G$ induces a family of weight systems $(w^t_\E)_t$ on $\G_\E$ indexed by $t\colon\E\to\R_{>0}$, as follows:
if $e$ is an edge of $\G$ which does not belong to $\E$, set $w^t_\E(e)=w(e)$; if $e\in\E$ is cut into two edges $e',e''$ of $\G_\E$, set
$w^t_\E(e')=t(e)w(e)^{1/2}$ and $w^t_\E(e'')=t(e)^{-1}w(e)^{1/2}$. Note that this family of weight systems is an orbit under the action
of the subgroup of ${\mathcal G}(\G_\E)$ consisting of elements $s$ such that $s(v)=1$ for all $v\in V(\G)$ and $s(v')=s(v'')$ whenever
$v',v''\in\partial\G_\E$ come from the same edge of $\E$.

Let us now formulate how the cutting affects the partition function. The proof is straightforward.

\begin{prop}\label{prop:cut-graph}
Fix a boundary condition $\partial D_0$ on $(\G,\partial\G)$ and a set $\E$ of edges of $\G$. Then,
given any parameter $t\colon\E\to\R_{>0}$,
\[
Z(\G;w\,|\,\partial D_0)=\sum_{I\subset\E} Z(\G_\E;w^t_\E\,|\,\partial D_0^I),
\]
where the sum is taken over all subsets $I$ of $\E$ and
$\partial D_0^I$ is the boundary condition on $(\G_\E,\partial\G_\E)$ induced by $\partial D_0$ and $I$:
a vertex of $\partial\G_\E$ is matched in $\partial D_0^I$
if and only if it is matched in $\partial D_0$ or it comes from an edge in $I$. \qed
\end{prop}

The operation opposite to cutting is called {\em gluing\/}: pick a pair of boundary vertices of $\G$, and glue the adjacent
edges $e',e''$ along these vertices into a single edge $e$. In order for the result to be a graph, it should be assumed that $e'$ and $e''$
are different edges of $\G$. We shall denote by $(\G_\varphi,\partial\G_\varphi)$ the graph obtained by gluing $(\G,\partial\G)$ according to a pairing
$\varphi$ of several vertices of $\partial\G$.

Note that a dimer configuration $D\in\D(\G,\partial\G)$ induces a dimer configuration $D_\varphi\in\D(\G_\varphi,\partial\G_\varphi)$ if and only if
the boundary condition $\partial D$ on $\partial\G$ is {\em compatible with $\varphi$\/}, i.e: $\varphi$ relates matched vertices with matched vertices.
Obviously, a dimer configuration $D_\E$ is compatible with the pairing $\varphi$ which glues back the edges of $\E$, and $(D_\E)_\varphi=D$
on $((\Gamma_\E)_\varphi,(\partial\Gamma_\E)_\varphi)=(\G,\partial\G)$.

An edge weight system $w$ on $\G$ induces an edge weight system $w_\varphi$ on $\G_\varphi$ as follows:
\[
w_\varphi(e)=\begin{cases}
w(e) & \text{if $e$ is an edge of $\G$;} \\
w(e')w(e'')& \text{if $e$ is obtained by gluing the edges $e'$ and $e''$ of $\G$.}
\end{cases}
\]
If $\E$ is a set of edges of $\G$ and $\varphi$ is the pairing which glues back these edges, then $(w_\E^t)_\varphi=w$ for any $t\colon\E\to\R_{>0}$.

The effect of gluing on the partition function is best understood in the language of quantum field theory.
We therefore postpone its study to Section \ref{section:QFT}.

\subsection{Cutting and gluing surface graphs with boundary}
\label{sub:glue-surface}

\begin{figure}[htbp]
\labellist\small\hair 2.5pt
\pinlabel {$\GSS$} at 0 240
\pinlabel {$C$} at 165 250
\pinlabel {$\G_C\subset\SI_C$} at 500 240 
\endlabellist
\centerline{\psfig{file=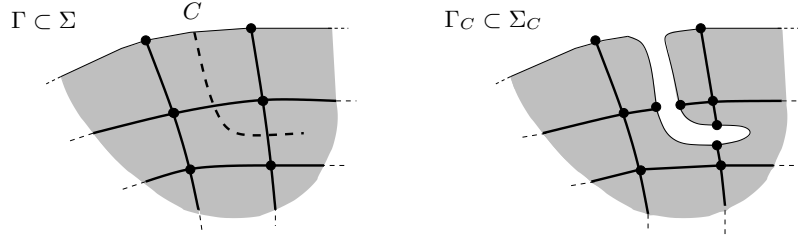,width=4in}}
\caption{Cutting a surface graph $\GSS$ along a curve $C$.}
\label{cut}
\end{figure}

Let $\G\subset \SI$ be a surface graph with boundary.
Let $C$ be a simple curve in $\SI$ which is ``in general position" with respect to $\G$, in the following sense:
\begin{romanlist}
\item{it is disjoint from the set of vertices of $\G$;}
\item{it intersects the edges of $\G$ transversally;}
\item{its intersection with any given face of $\SI$ is connected.}
\end{romanlist}
Let $\SI_C$ be the surface with boundary obtained by cutting $\SI$ open along $C$.
Also, let $\G_C:=\G_{\E(C)}$ be the graph with boundary obtained by cutting $(\G,\partial\G)$ along the set $\E(C)$ of edges of $\G$
which intersect $C$, as illustrated in Figure \ref{cut}.

\medskip

Obviously, $\G_C\subset \SI_C$ is a surface graph with boundary.
We will say that it is obtained by {\em cutting $\G\subset\SI$ along $C$}.
Abusing notation, we shall write $w^t_C$ for the weight system $w^t_{\E(C)}$ on $\G_C$.

A class $\beta\in H_1(\SI,\partial \SI; \Z_2)$ induces $\beta_C\in H_1(\SI_C,\partial\SI_C;\Z_2)$ via
\[
H_1(\SI, \partial \SI; \Z_2)\to H_1(\SI, \partial \SI\cup N(C);
\Z_2)\simeq H_1(\SI_C, \partial \SI_C; \Z_2).
\]
Here $N(C)$ denotes a neighborhood of $C$ in $\SI$, the first
homomorphism is induced by inclusion, and the second one is the
excision isomorphism. Note that given any two dimers configurations $D$ and $D'$ on $\GSS$,
$\Delta(D_C,D'_C)=\Delta(D,D')_C$ in $H_1(\SI_C, \partial \SI_C; \Z_2)$.

This easily leads to the following refinement of Proposition \ref{prop:cut-graph}.

\begin{prop}\label{prop:partition}
Fix $\beta \in H_1(\SI, \partial \SI; \Z_2)$, $D'\in \D(\G,\partial \G)$, and a boundary condition $\partial D_0$ on
$(\G,\partial \G)$. Then, given any parameter $t\colon\E(C)\to\R_{>0}$,
\[
Z_{\beta, D'}(\G;w\,|\,\partial D_0)=\sum_{I\subset\E(C)} Z_{\beta_C,D'_C}(\G_C;w^t_C\,|\,\partial D_0^I),
\]
where the sum is taken over all subsets $I$ of $\E(C)$ and $\partial D_0^I$ is the boundary condition on $(\G_C,\partial\G_C)$
induced by $\partial D_0$ and $I$.\qed
\end{prop}

\begin{figure}[htbp]
\labellist\small\hair 2.5pt
\pinlabel {$\GSS$} at 200 420
\pinlabel {$\G_\varphi\subset\SI_\varphi$} at 800 420 
\pinlabel {$\scriptstyle{M_1\;\stackrel{\varphi}{\longrightarrow}\;M_2}$} at 205 280 
\endlabellist
\centerline{\psfig{file=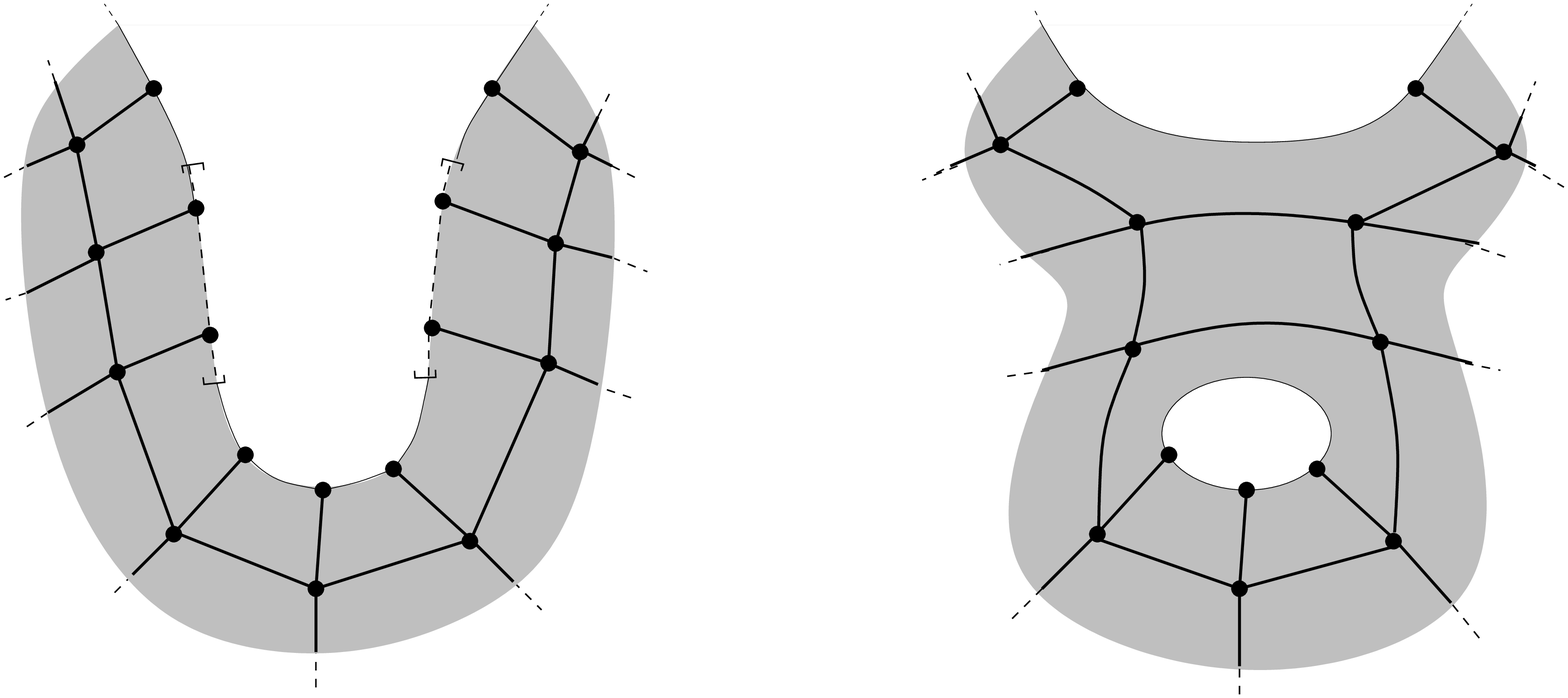,width=4in}}
\caption{Gluing a surface graph $\GSS$ along $\varphi\colon M_1\to M_2$.}
\label{fig:glue}
\end{figure}

Let us now define the operation opposite to cutting a surface graph with boundary. 
Pick two closed connected subsets $M_1,M_2$ of $\partial\SI$, which are not points, and satisfy the following properties:
\begin{romanlist}
\item{$M_1\cap M_2\subset\partial M_1\cup\partial M_2$ and $\partial M_1\cup\partial M_2$ is disjoint from $\partial\G$;}
\item{the intersection of each given face of $\SI$ with $M_1\cup M_2$ is connected;}
\item{there exists an orientation-reversing homeomorphism $\varphi\colon M_1\to M_2$ which induces a bijection $M_1\cap\partial\G\to M_2\cap\partial\G$
such that for all $v$ in $M_1\cap\partial\G$, $v$ and $\varphi(v)$ are not adjacent to the same edge of $\G$.}
\end{romanlist}
Let $\G_\varphi\subset\SI_\varphi$ be obtained from the surface graph $\GSS$ by identifying $M_1$ and $M_2$ via $\varphi$ and removing
the corresponding vertices of $\Gamma$. This is illustrated in Figure \ref{fig:glue}.
By the conditions above, the pair $\G_\varphi\subset\SI_\varphi$ remains a surface graph. It is said to be obtained by
{\em gluing $\GSS$ along $\varphi$.}

Note that any surface graph $\G_C\subset\SI_C$ obtained by cutting $\GSS$ along some curve $C$ in general position
with respect to $\G$ satisfies the conditions
listed above. Furthermore, $(\G_C)_\varphi\subset(\SI_C)_\varphi=\GSS$, where $\varphi$ is the obvious homeomorphism identifying the two closed subsets
of $\partial\SI_C$ coming from $C$. Conversely, if $C$ denotes the curve in $\SI_\varphi$ given by the identification of $M_1$ and $M_2$ via $\varphi$,
then it is in general position with respect to $\G_\varphi$, and $(\G_\varphi)_C\subset(\SI_\varphi)_C$.

\subsection{Cutting and gluing discrete spin structures}
\label{sub:glue-spin}

Let $\G\subset \SI$ be a surface graph with boundary, and let $C$
be a simple curve in $\SI$ in general position with respect to $\G$. As noted above,
any dimer configuration $D$ on $(\G, \partial\G)$ induces a dimer configuration
$D_C$ on $(\G_C,\partial\G_C)$. If two dimer configurations $D, D'\in\D(\G,\partial\G)$ are
equivalent, then $D_C, D_C'\in\D(\G_C, \partial \G_C)$ are equivalent as well:
\[
\Delta(D_C, D_C')=\Delta(D, D')_C=0 \in H_1(\SI_C, \partial \SI_C; \Z_2).
\]

A Kasteleyn orientation $K$ on $\G\subset\SI$ induces a Kasteleyn orientation $K_C$ on $\G_C\subset\SI_C$ as follows.
Let $K_C$ be equal to $K$ on all edges of $\overline\G_C$ coming from edges of $\overline\G$.
For all the new edges of $\overline\G_C$, there is a unique orientation which satisfies the Kasteleyn condition, since each face of
$\SI$ is crossed at most once by $C$.
One easily checks that if $K$ and $K'$ are equivalent Kasteleyn orientations, then $K_C$ and $K'_C$ are also equivalent.
Hence, there is a well-defined operation of cutting discrete spin structures on a surface
with boundary.

This is not a surprise. Indeed, the inclusion $\SI_C\subset\SI_C\cup N(C)=\SI$ induces a homomorphism
$i_*\colon H_1(\SI_C;\Z_2)\to H_1(\SI;\Z_2)$. The assignment $q\mapsto q_C=q\circ i_*$ defines a map from the quadratic forms
on $ H_1(\SI;\Z_2)$ to the quadratic forms on $H_1(\SI_C;\Z_2)$, which is affine over the restriction homomorphism
$i^*\colon H^1(\SI;\Z_2)\to H^1(\SI_C;\Z_2)$.
By Johnson's theorem, it induces an affine map between the sets of spin structures ${\mathcal S}(\SI)\to{\mathcal S}(\SI_C)$.
By Corollary \ref{cor:spin}, there is a unique map $\K(\GSS)\to\K(\Gamma_C\subset\SI_C)$ which makes the following diagram commute:
\begin{equation}\label{equ:dia}
\xymatrix{
\K(\GSS)\ar[d]_\cong^{\psi_D}\ar[r]& \K(\Gamma_C\subset\SI_C)\ar[d]_\cong^{\psi_{D_C}}\\
{\mathcal S}(\SI)\ar[r]& {\mathcal S}(\SI_C).
}
\end{equation}
This map is nothing but $[K]\mapsto[K_C]$.

\medskip

Now, let $K$ be a Kasteleyn orientation on a surface graph $\GSS$, and let $\varphi\colon M_1\to M_2$ be an orientation-reversing homeomorphism
between two closed connected subsets in $\partial\SI$, as described above. We shall say that a Kasteleyn orientation $K$ on $\GSS$ is
{\em compatible with $\varphi$\/} if the following conditions hold:
\begin{romanlist}
\item{whenever two edges $e',e''$ of $\overline\G$ are glued into a single edge $e$ of $\overline\G_\varphi$, the orientation $K$ agrees on $e'$ and $e''$,
giving an orientation $K_\varphi$ on $e$;}
\item{the induced orientation $K_\varphi$ is a Kasteleyn orientation on $\G_\varphi\subset\SI_\varphi$.}
\end{romanlist}
The Kasteleyn orientation $K_\varphi$ on $\G_\varphi\subset\SI_\varphi$ is said to be obtained by {\em gluing $K$ along $\varphi$\/}.

Given any Kasteleyn orientation $K$ on $\GSS$, the induced orientation $K_C$ on $\G_C\subset\SI_C$
is compatible with the map $\varphi$ such that $(\SI_C)_\varphi=\SI$; furthermore, $(K_C)_\varphi$ is equal to $K$.
Conversely, if $K$ is a Kasteleyn orientation on $\GSS$ which is  compatible with $\varphi$, and $C$ denotes the curve in $\SI_\varphi$ given by the identification of $M_1$ and $M_2$ via $\varphi$, then $(K_\varphi)_C$ is equal to $K$. With these notations, any dimer configuration $D$ on $\G$
which is compatible with $\varphi$ satisfies $(D_\varphi)_C=D$. Therefore, diagram (\ref{equ:dia}) gives
\[
\xymatrix{
\K(\G_\varphi\subset\SI_\varphi)\ar[d]_\cong^{\psi_{D_\varphi}}\ar[r]& \K(\GSS)\ar[d]_\cong^{\psi_D}\\
{\mathcal S}(\SI_\varphi)\ar[r]& {\mathcal S}(\SI),
}
\]
where both horizontal maps are affine over $i^*\colon H^1(\SI_\varphi;\Z_2)\to H^1(\SI;\Z_2)$. Understanding the gluing of Kasteleyn orientations
(up to equivalence) now amounts to understanding the restriction homomorphism $i^*$. Using the exact sequence of the pair $(\SI_\varphi,\SI)$, one
easily checks the following results:
\begin{ticklist}
\item{The restriction homomorphism $i^*$ is injective, unless $M_1$ and $M_2$ are disjoint and belong to the same connected component of $\SI$.
In this case, the kernel of $i^*$ has dimension 1.}
\item{The homomorphism $i^*$ is onto unless $M_1\cup M_2$ is a 1-cycle and the corresponding connected component
of $\SI_\varphi$ is not closed. In this case, the cokernel of $i^*$ has dimension 1.}
\end{ticklist}
This leads to the four following cases. Fix a Kasteleyn orientation $K$ on $\GSS$.
\begin{enumerate}
\item{If $i^*$ is an isomorphism, then there exist a Kasteleyn orientation $K'$ equivalent to $K$ which is compatible with $\varphi$.
Furthermore, the assignment
$[K]\mapsto[K'_\varphi]$ gives a well-defined map between $\K(\GSS)$ and $\K(\G_\varphi\subset\SI_\varphi)$.}
\item{If $i^*$ is onto but not injective, then there exist $K',K''\sim K$ which
are compatible with $\varphi$, inducing two distinct well-defined maps $[K]\mapsto[K'_\varphi]$ and $[K]\mapsto[K''_\varphi]$ between $\K(\GSS)$ and
$\K(\G_\varphi\subset\SI_\varphi)$.}
\item{If $i^*$ is injective but not onto, then $M_1\cup M_2$ is a 1-cycle, oriented as part of the boundary of $\SI$. There exist $K'\sim K$
which is compatible with $\varphi$ if and only if the following condition holds:
\begin{align*}
n^K(M_1)+n^K(M_2)&\equiv 0\pmod{2}\quad\text{if $M_1$ and $M_2$ are disjoint;}\\
n^K(M_1\cup M_2)&\equiv 1\pmod{2}\quad\text{otherwise.}
\end{align*}
(Note that this condition only depends on the equivalence class of $K$.) In this case, it induces a well-defined class
$[K'_\varphi]$ in $\K(\G_\varphi\subset\SI_\varphi)$.}
\item{Finally, assume $i^*$ is neither onto nor injective. If $K$ satisfies the condition above, then there exist $K',K''\sim K$ which
are compatible with $\varphi$, inducing two well-defined maps $[K]\mapsto[K'_\varphi]$ and $[K]\mapsto[K''_\varphi]$. On the other hand,
if $K$ does not satisfy the condition above, then it does
not contain any representative which is compatible with $\varphi$.}
\end{enumerate}

\subsection{Cutting Pfaffians}
\label{sub:cut-Pfaffian}

Let us conclude this section with one last observation. Let $\GSS$ be a surface graph with boundary, and let $C$ be a simple curve in $\SI$.
The equality
\[
Z(\G;w\,|\,\partial D_0)=\sum_{I\subset\E(C)} Z(\G_C;w^t_C\,|\,\partial D_0^I)
\]
of Proposition \ref{prop:cut-graph} can be understood as the Taylor series expansion of the function $Z(\G;w\,|\,\partial D_0)$ in the variables
$(w(e))_{e\in\E(C)}$. Clearly, if $\E(C)=\{e_{i_1},\dots,e_{i_k}\}$, then
\[
\prod_{\ell=1}^kw(e_{i_\ell})\frac{\partial^kZ(\G;w\,|\,\partial D_0)}{\partial w(e_{i_1})\cdots\partial w(e_{i_k})}(0)=Z(\G_C;w^t_C\,|\,\partial D_0^I).
\]
By Theorem \ref{thm:Pfaffian}, the partition function $Z(\G;w\,|\,\partial D_0)$ can be expressed as a linear combination of Pfaffians of
matrices $A^K(\overline\G;w\,|\,\partial D_0)$ depending on Kasteleyn orientations $K$ of $\GSS$ such that $q^K_{D_0}(\gamma)=0$ for all boundary component
$\gamma$ of $\SI$. Recall that any such orientation $K$ extends to a Kasteleyn orientation $K_C$ on $\G_C\subset\SI_C$. Furthermore, all equivalence
classes of Kasteleyn orientations such that $q^{K_C}_{(D_0)_C}(\gamma)=0$ for all boundary component $\gamma$ of $\SI_C$
are obtained in this way. (This follows from the fact that the map $[K]\mapsto[K_C]$ is affine over the restriction homomorphism.)
Finally, the partition function $Z(\G_C;w^t_C\,|\,\partial D_0)$ can also be expressed as a linear combination of Pfaffians
of matrices $A^{K_C}(\overline\G_C;w^t_C\,|\,\partial D_0)$ via Theorem \ref{thm:Pfaffian}.

Gathering all these equations, we obtain a relation between the Pfaffian of the matrix $A^K(\overline\G;w\,|\,\partial D_0)$ and the Pfaffian of the
matrix $A^{K_C}(\overline\G_C;w^t_C\,|\,\partial D_0)$. This relation turns out to be exactly the equation below, a well-known property of Pfaffians.

\begin{prop}
Let $A=(a_{ij})$ be a skew-symmetric matrix of size $2n$. Given an ordered subset $I$ of the ordered set $\alpha=(1,\dots,2n)$,
let $A_I$ denote the matrix obtained from $A$ by removing the $i^\mathit{th}$ row and the $i^\mathit{th}$ column for all $i\in I$.
Then, for any ordered set of indices $I=(i_1,j_1,\dots,i_k,j_k)$,
\[
\frac{\partial^k\Pf(A)}{\partial a_{i_1j_1}\cdots\partial a_{i_kj_k}}=(-1)^{\sigma(I)}\Pf(A_I),
\]
where $(-1)^{\sigma(I)}$ denote the signature of the permutation which sends $\alpha$ to the ordered set $I(\alpha\backslash I)$.\qed
\end{prop}

\section{Quantum field theory for dimers}
\label{section:QFT}

\subsection{Quantum field theory on graphs}
\label{sub:graphQFT}

Let $(\G,\partial\G)$ be a graph with boundary, and let us assume that each vertex $v$ in $\partial\G$ is oriented, that is,
endowed with some sign $\e_v$. In the spirit of the Atiyah-Segal axioms for a $(0+1)$-topological quantum field theory \cite{At,Segal},
let us define a {\em quantum field theory on graphs\/} as the following assignment:
\begin{enumerate}
\item{Fix a finite dimensional complex vector space $V$.}
\item{To the oriented boundary $\partial\G$, assign the vector space
\[
Z(\partial\G)=\bigotimes_{\genfrac{}{}{0pt}{}{v\in\partial\G}{\e_v=+1}} V\otimes\bigotimes_{\genfrac{}{}{0pt}{}{v\in\partial\G}{\e_v=-1}} V^*,
\]
where $V^*$ denotes the vector space dual to $V$.}
\item{To a finite graph $\G$ with oriented boundary $\partial\G$ and weight system $w$, assign some vector $Z(\G;w)\in Z(\partial\G)$,
with $Z(\emptyset;w)=1\in\C=Z(\emptyset)$.}
\end{enumerate}

Note that any orientation preserving bijection $f\colon\partial\G\to\partial\G'$ induces an isomorphism
$Z(f)\colon Z(\partial\G)\to Z(\partial\G')$ given by permutation of the factors. This assignment is functorial: if $g\colon\partial\G'\to\partial\G''$
is another orientation preserving bijection, then $Z(g\circ f)=Z(g)\circ Z(f)$. Finally, if $f\colon\partial\G\to\partial\G'$ extends to a
homeomorphism $F\colon\G\to\G'$, then $Z(f)$ maps $Z(\G)$ to $Z(\G')$.
Note also that $Z(-\partial\G)=Z(\partial\G)^*$, and that $Z(\partial\G\sqcup\partial\G')=Z(\partial\G)\otimes Z(\partial\G')$.

The main point is that we require the following {\em gluing axiom\/}.
Let $\G$ be a graph with oriented boundary $\partial\G$, such that there exists two disjoint subsets
$X_1,X_2$ of $\partial\G$ and an orientation reversing bijection $\varphi\colon X_1\to X_2$ (i.e. $\e_{\varphi(v)}=-\e_v$ for all $v\in X_1$).
Obviously, $\varphi$ induces a linear isomorphism $Z(\varphi)\colon Z(X_1)\to Z(X_2)^*$.
Let $\G_\varphi$ denote the graph with boundary $\partial\G_\varphi=\partial\G\setminus(X_1\cup X_2)$ obtained by gluing $\Gamma$ according to
$\varphi$, and let $w_\varphi$ be the corresponding weight system on $\G_\varphi$ (recall Section \ref{sub:glue-graph}). Let $B_\varphi$ denote the
composition
\[
Z(\partial\G)=Z(\partial\G_\varphi)\otimes Z(X_1)\otimes Z(X_2)\to
Z(\partial\G_\varphi)\otimes Z(X_2)^*\otimes Z(X_2)\to Z(\partial\G_\varphi),
\]
where the first homomorphism is given by $id\otimes Z(\varphi)\otimes id$, and the second is induced by the natural pairing
$Z(X_2)^*\otimes Z(X_2)\to\C$. We require that
\[
B_\varphi(Z(\G;w))=Z(\G_\varphi;w_\varphi).
\]
\begin{rem}
In the same spirit, one can define a {\em quantum field theory on surface graphs\/}. Here, the vector $Z(\GSS;w)\in Z(\partial\G)$
might depend on the realization of $\G$ as a surface graph $\GSS$, and the gluing axiom concerns gluing of surface graphs, as defined in Section
\ref{sub:glue-surface}.
\end{rem}

\subsection{Quantum field theory for dimers on graphs}
\label{sub:dimerQFT}

Let us now explain how the dimer model on weighted graphs with boundary defines a quantum field theory.
As vector space $V$, choose the 2-dimensional complex vector space with fixed basis $a_0,a_1$. Let $\alpha_0,\alpha_1$ denote the dual basis in $V^*$.
To a finite graph $\G$ with oriented boundary $\partial\G$ and weight system $w$, assign
\[
Z(\G;w)=\sum_{\partial D}Z(\G;w\,|\,\partial D)\,a(\partial D)\in Z(\partial\G),
\]
where the sum is on all possible boundary conditions $\partial D$ on $\partial\G$, and
\[
a(\partial D)=
\bigotimes_{\genfrac{}{}{0pt}{}{v\in\partial\G}{\e_v=+1}}a_{i_v(\partial D)}\otimes
\bigotimes_{\genfrac{}{}{0pt}{}{v\in\partial\G}{\e_v=-1}}\alpha_{i_v(\partial D)}\in Z(\partial\G).
\]
Here, $i_v(\partial D)=1$ if the vertex $v$ is matched by $\partial D$, and $i_v(\partial D)=0$ otherwise.

Let us check the gluing axiom. First note that $B_\varphi(a(\partial D))=0$ unless $\partial D$ is compatible with $\varphi$ 
(i.e: unless $\varphi(v)$ is matched in $\partial D$ if and only if $v$ is matched in $\partial D$). In such a case,
$B(a(\partial D))=a(\partial D|_{\partial\G_\varphi})$, where $\partial D|_{\partial\G_\varphi}$ denotes the restriction of the boundary condition
$\partial D$ to $\partial\G_\varphi\subset\partial\G$. All the possible boundary conditions $\partial D_\varphi$ on $\partial\G_\varphi$ are given
by such restrictions. Therefore,
\[
B_\varphi(Z(\G;w))=\sum_{\partial D_\varphi}\Big(\sum_{\partial D\supset\partial D_\varphi}Z(\G;w\,|\,\partial D)\Big)a(\partial D_\varphi),
\]
the interior sum being on all boundary conditions $\partial D$ on $\partial\G$ that are compatible with $\varphi$, and such that
$\partial D|_{\partial\G_\varphi}=\partial D_\varphi$. By definition,
\[
\sum_{\partial D\supset\partial D_\varphi}Z(\G;w\,|\,\partial D)=\sum_{D:\partial D\supset\partial D_\varphi}w(D)
=Z(\G_\varphi;w_\varphi\,|\,\partial D_\varphi).
\]
Therefore, the gluing axiom is satisfied.

\subsection{The dimer model as the theory of free Fermions}
\label{sub:FermionsQFT}

Let $W$ be an $n$-dimensional vector space. The choice of an ordered basis in $W$ induces an isomorphism between its exterior algebra
$\bigwedge W=\oplus_{k=0}^n\bigwedge^k W$ and the algebra generated by elements $\phi_1,\dots,\phi_n$ with defining relations
$\phi_i\phi_j=-\phi_j\phi_i$. This space is known as the {\em Grassman algebra\/} generated by $\phi_1,\dots, \phi_n$.
The choice of an ordered basis in $W$ also defines a basis in the top exterior
power of $W$. The {\em integral over the Grassman algebra of $W$\/} of an element $a\in\bigwedge W$
is the coordinate of $a$ in the top exterior power of $W$ with respect to this basis. It is denoted by $\int a\,d\phi$.

There is a scalar product on the Grassman algebra generated by $\phi_1,\dots,\phi_n$; it is given by the Grassman integral
\begin{equation}\label{equ:scalar}
<F,G>=\int \exp\Big(\sum_{i=1}^n\phi_i\psi_i\Big)F(\phi)G(\psi)d\phi d\psi.
\end{equation}
Note that the monomial basis is orthonormal with respect to this scalar product.
One easily shows (see e.g. the Appendix to \cite{C-R}) that the Pfaffian of a skew symmetric matrix $A=(a_{ij})$ can be written as
\[
\Pf(A)=\int \exp\Big(\frac{1}{2}\sum_{i,j=1}^n\phi_i a_{ij}\phi_j\Big)d\phi.
\]

Let us now use this to reformulate the quantum field theory of dimers in terms of Grassman integrals.
Let $\GSS$ be a (possibly disconnected) surface graph, possibly with boundary.
Let us fix a numbering of the vertices of $\G$, a boundary condition $\partial D_0$ on
$\partial\G$ and a Kasteleyn orientation $K$ on $\GSS$. Let $a^K_{ij}$ be the Kasteleyn coefficient associated to $K$ and the vertices $i,j$ of $\G$
(recall Section \ref{section:Kasteleyn}). By Theorem \ref{thm:Pfaffian} and the identity above,
\[
Z(\Gamma;w\,|\,\partial D_0)=\frac{1}{2^g}\sum_{[K]}\A(q_{D_0}^K)\e^K(D_0)
\int\exp\Big(\frac{1}{2}\sum_{i,j\in V(D_0)}\phi_i a^K_{ij} \phi_j\Big)d\phi_{\partial D_0},
\]
where the sum is over all $2^{b_1(\SI)}$ equivalence classes of Kasteleyn orientations on $\GSS$,
$V(D_0)$ denotes the set of vertices of $\G$ that are matched by $D_0$, and $d\phi_{\partial D_0}=\wedge_{i\in V(D_0)}d\phi_i$.
This leads to the formula
\[
Z(\Gamma;w\,|\,\partial D_0)=\frac{1}{2^{g}}\sum_{[K]}\int\exp\Big(\frac{1}{2}\sum_{i,j\in V(D_0)}\phi_i a^K_{ij} \phi_j\Big)D_{\partial D_0}^K\phi,
\]
where $D_{\partial D_0}^K\phi=\A(q^{K}_{D_0})\e^{K}(D_0)\,d_{\partial D_0}\phi$.
Let us point out that this measure does not depend on the choice of $D_0$, but only on the induced boundary condition $\partial D_0$.

Now, the numbering of the vertices of $\G$ gives a numbering of the vertices of $\partial\G$. This induces a linear isomorphism
between $Z(\partial\G)$ and the Grassman algebra $\bigwedge(\partial\G)$ generated by $(\phi_i)_{i\in\partial\G}$.
The image of the partition function under this isomorphism is the following element of the Grassman algebra of boundary vertices:
\[
Z(\G;w)=\sum_{\partial D_0}Z(\G;w\,|\,\partial D_0)\,\textstyle\prod_{i\in V(\partial D_0)}\phi_i\,\in\,\bigwedge(\partial\G),
\]
where $V(\partial D_0)=V(D_0)\cap\partial\G$. This leads to
\begin{align*}
Z(\G;w)&=\frac{1}{2^{g}}\sum_{\partial D_0}\sum_{[K]}\int\exp\Big(\frac{1}{2}\sum_{i,j\in V(D_0)}\phi_i a^K_{ij} \phi_j\Big)D_{\partial D_0}^K\phi
\,\textstyle\prod_{i\in V(\partial D_0)}\phi_i\\
&=\frac{1}{2^{g}}\sum_{[K]}\int\exp\Big(\frac{1}{2}\sum_{i,j\in V(\G)}\phi_i a^K_{ij} \phi_j\Big)D^K\phi,
\end{align*}
where $D^K\phi=\A(q^{K}_{D_0})\e^{K}(D_0)\wedge_{i\notin\partial\G}d\phi_i$. This measure depends only on $K$, but not on $D_0$.

We can now formulate the dimer model as the theory of free (Gaussian) Fermions:
\begin{enumerate}
\item{To the boundary of $\Gamma\subset\Sigma$, we assign $\bigwedge(\partial\G)$,
the Grassman algebra generated by the ordered set $\partial\G$;}
\item{To a surface graph $\GSS$ with ordered set of vertices $V(\G)$ and weight system $w$,
we assign the element $Z(\G\subset\SI;w)$ of $\bigwedge(\partial\G)$ given by
\[
Z(\G\subset\SI;w)=\frac{1}{2^{g}}\sum_{[K]}\int\exp\Big(\frac{1}{2}\sum_{i,j\in V(\G)}\phi_i a^K_{ij} \phi_j\Big)D^K\phi,
\]
where the sum is over all $2^{b_1(\SI)}$ equivalence classes of Kasteleyn orientations on $\GSS$, and
$D^K\phi=\A(q^{K}_{D_0})\e^{K}(D_0)\wedge_{i\notin\partial\G}d\phi_i$.}
\end{enumerate}

The gluing axiom now takes the following form. Let $\G_\varphi\subset\SI_\varphi$ denote the surface graph with boundary obtained by gluing
$\GSS$ along some orientation-reversing homeomorphism $\varphi\colon M_1\to M_2$ (see Section \ref{sub:glue-surface}). Recall that $\varphi$ induces
a bijection between the two disjoint sets $X_1=\partial\G\cap M_1$ and $X_2=\partial\G\cap M_2$. Therefore, it induces an isomorphism
$Z(\varphi)\colon\bigwedge(X_1)\to\bigwedge(X_2)$. Consider the map $B_\varphi$ given by the composition
\[
\textstyle
\bigwedge(\partial\G)=\bigwedge(\partial\G_\varphi)\otimes \bigwedge(X_1)\otimes \bigwedge(X_2)\to
\bigwedge(\partial\G_\varphi)\otimes \bigwedge(X_2)^*\otimes \bigwedge(X_2)\to \bigwedge(\partial\G_\varphi).
\]
Here, the first homomorphism is given by $id\otimes (h\circ Z(\varphi))\otimes id$, where $h\colon\bigwedge(X_2)\to\bigwedge(X_2)^*$
is the isomorphism induced by the scalar product (\ref{equ:scalar}). Then, we require that
\[
B_\varphi(Z(\GSS;w))=Z(\G_\varphi\subset\SI_\varphi;w_\varphi).
\]

We already know that this equality holds. Indeed, $Z(\GSS;w)$ just depends on $(\G,w)$, and
the formula above is nothing but the gluing axiom for $Z(\G;w)$ translated in the formalism of Grassman
algebras. However, it can also be proved from scratch using the results of Section \ref{sub:glue-spin} together with well-known
properties of Pfaffians.

\section{Dimers on bipartite graphs and height functions}
\label{section:bipartite}

\subsection{Composition cycles on bipartite graphs}
\label{sub:bip-cc}

Recall that a {\em bipartite structure\/} on a graph $\G$ is a partition of its set of vertices into two groups,
say blacks and whites, such that no edge of $\G$ joins two vertices of the same group. Equivalently, a bipartite
structure can be regarded as a 0-chain
\[
\beta =\sum_{v\text{ black}} v-\sum_{v\text{ white}} v\;\in C_0(\G;\Z).
\]

A bipartite structure induces an orientation on the edges of $\G$, called the {\em bipartite orientation\/}: simply
orient all the edges from the white vertices to the black ones. Using this orientation, a dimer configuration $D\in\D(\G,\partial\G)$ can now be regarded
as a 1-chain with $\Z$-coefficients
\[
D=\sum_{e\in D}e\,\in C_1(\G;\Z)
\]
such that $\partial D=\beta$ in $C_0(\G,\partial\G;\Z)=C_0(\G;\Z)/C_0(\partial\G;\Z)$.
Therefore, given two dimer configurations $D,D'$ on $\G$, their difference $D-D'$ is a 1-cycle $(rel\;\partial\G)$
with $\Z$-coefficients, denoted by $C(D,D')$. Its connected components are called {\em $(D,D')$-composition cycles\/}.
In short, a bipartite structure on a graph allows to orient the composition cycles.

\subsection{Height functions for planar bipartite graphs}
\label{sub:bi-h}

\begin{figure}[htbp]
\labellist\small\hair 2.5pt
\pinlabel {$0$} at 400 0
\pinlabel {$0$} at 460 60
\pinlabel {$0$} at 500 125 
\pinlabel {$0$} at 550 195 
\pinlabel {$1$} at 580 105
\pinlabel {$-1$} at 436 200 
\endlabellist
\centerline{\psfig{file=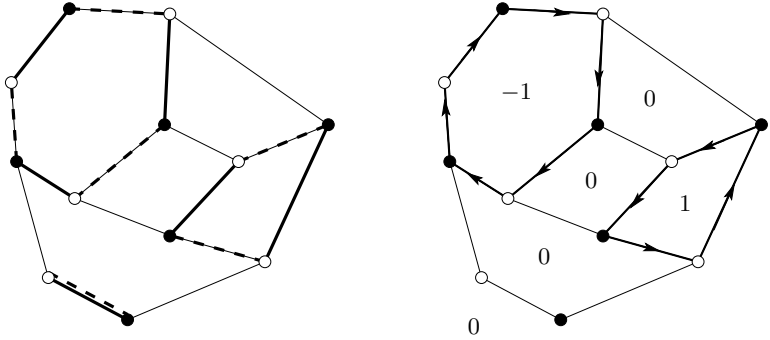,width=4in}}
\caption{An example of a bipartite planar graph with two dimer configurations $D$ (solid) and $D'$ (traced lines). The corresponding height function
$h_{D,D'}$ (where $f_0$ is the outer face) and $(D,D')$-composition cycles are pictured on the right hand side.}
\label{fig:height}
\end{figure}

Let us now assume that the bipartite graph $\G$ is planar without boundary, i.e. that it can be realized as a surface graph $\G\subset S^2$.
Let $X$ denote the induced
cellular decomposition of the 2-sphere, which we endow with the counter-clockwise orientation.
Since $H_1(X;\Z)=H_1(S^2;\Z)=0$, the 1-cycle $C(D,D')$ is a 1-boundary, so there exists $\sigma_{D,D'}\in C_2(X;\Z)$ such that
$\partial\sigma_{D,D'}=C(D,D')$. Let $h_{D,D'}\in C^2(X;\Z)$ be given by the equality
\[
\sigma_{D,D'}=\sum_{f\in F(X)}h_{D,D'}(f)\,f\in C_2(X;\Z),
\]
where the sum is over all faces of $X$. The cellular 2-cochain $h_{D,D'}$ is called a {\em height function associated to $D,D'$\/}.
Since $H_2(X;\Z)=H_2(S^2;\Z)=\Z$, the 2-chain $\sigma_{D,D'}$ is uniquely defined by $D,D'$ up to a constant, and the same holds for $h_{D,D'}$.
Hence, one can normalize all height functions by setting $h_{D,D'}(f_0)=0$ for some fixed face $f_0$. This is illustrated in Figure \ref{fig:height}.

Alternatively, $h_{D,D'}$ can be defined as the only $h\in C^2(X;\Z)$ such that $h(f_0)=0$ and $h$ increases by 1 when a $(D,D')$-composition
cycle is crossed in the positive direction (left to right as we cross). It follows that for any height function $h$ and any two $2$-cells $f_1$ and
$f_2$,
\[
|h(f_1)-h(f_2)|\leq d(f_1,f_2),
\]
where $d(f_1,f_2)$ is the distance between $f_1$ and $f_2$ in the dual graph, i.e. the
minimal number of edges crossed by a path connecting an point inside $f_1$ with a point inside $f_2$. This can be regarded as a
Lipschitz property of height functions.
Note also that for any three dimer configurations $D$, $D'$ and $D''$ on $\G$, the following cocycle equality holds:
\[
h_{D,D'}+h_{D',D''}=h_{D,D''}.
\]

The Lipschitz condition stated above leads to the following definition. Given a fixed 2-cell $f_0$ of the cellular decomposition $X$ induced by
$\G\subset S^2$, set
\[
\mathcal{H}(X,f_0)=\{h\in C^2(X;\Z)\,|\,\text{$h(f_0)=0$ and $|h(f_1)-h(f_2)|\leq d(f_1,f_2)\;\forall f_1,f_2$}\}.
\]

Given $h\in \mathcal{H}(X,f_0)$, let $C(h)$ denote the oriented
closed curves formed by the set of oriented edges $e$ of $\G$ such that $h$ increases its value by 1 when crossing $e$ in the positive direction.
(In other words, $C(h)=\partial\sigma$, where $\sigma\in C_2(X;\Z)$ is dual to $h\in C^2(X;\Z)$.) Obviously, there is a well-defined map
\[
\D(\G)\times\D(\G)\to\mathcal{H}(X,f_0),\quad(D,D')\mapsto h_{D,D'}
\]
with $C(h_{D,D'})=C(D,D')$. However, this map is neither injective nor surjective in general. Indeed, the number of preimages of a given $h$ is equal to
the number of dimer configurations on the graph obtained from $\G$ by removing the star of $C(h)$. Depending on $\G\subset S^2$,
this number can be zero, or arbitrarily large.

To obtain a bijection, we proceed as follows. Fix a dimer configuration $D_0$ on $\G$.
Let ${\mathcal C}(D_0)$ denote the set of all $C\subset\G$ consisting of disjoint oriented simple 1-cycles, such that the following condition holds:
for all $e\in D_0$, either $e$ is contained in $C$ or $e$ is disjoint from $C$. Finally, set
\[
{\mathcal H}_{D_0}(X,f_0)=\{h\in{\mathcal H}(X,f_0)\,|\, C(h)\in{\mathcal C}(D_0)\}.
\]

\begin{prop}\label{prop:bij}
Given any $h\in\mathcal{H}_{D_0}(X,f_0)$, there is unique dimer configuration $D\in\D(\G)$ such that $h_{D,D_0}=h$. Furthermore,
given any two dimer configurations $D_0,D_1$ on $\G$, we have a canonical bijection
\[
\mathcal{H}_{D_0}(X,f_0)\to\mathcal{H}_{D_1}(X,f_0)
\]
given by $h\mapsto h+h_{D_0,D_1}$.
\end{prop}
\begin{proof}
One easily checks that the assignment $D\mapsto C(D,D_0)$ defines a bijection $\D(\G)\to {\mathcal C}(D_0)$. Furthermore, there is
an obvious bijection $\mathcal{H}_{D_0}(X,f_0)\to{\mathcal C}(D_0)$ given by $h\mapsto C(h)$. This induces a bijection $\D(\G)\to\mathcal{H}_{D_0}(X,f_0)$
and proves the first part of the proposition. The second part follows from the first one via the cocycle identity $h_{D,D_0}+h_{D_0,D_1}=h_{D,D_1}$.
\end{proof}

Let us now consider an edge weight system $w$ on the bipartite planar graph $\G$. Recall that the Gibbs measure of $D\in\D(\G)$
is given by
\[
\mbox{Prob}(D)=\frac{w(D)}{Z(\G; w)},
\]
where $w(D)=\prod_{e\in D} w(D)$ and $Z(\G; w)=\sum_{D\in\D(\G)} w(D)$. Let us now fix a dimer
configuration $D_0$ and a face $f_0$ of $X$, and use the bijection $\D(\G)\to\mathcal{H}_{D_0}(X,f_0)$ given by $D\mapsto h_{D,D_0}$ to translate
this measure into a probability measure on $\mathcal{H}_{D_0}(X,f_0)$.

To do so, we shall need the following notations: given an oriented edge $e$ of $\G$, set
\[
w_\beta(e)=
\begin{cases}
w(e) & \text{if the orientation on $e$ agrees with the bipartite orientation;} \\
w(e)^{-1} & \text{otherwise.}
\end{cases}
\]
This defines a group homomorphism $w_\beta\colon C_1(X;\Z)\to\R_{>0}$. Finally, given any $f\in F(X)$, set
\[
q_f=w_\beta(\partial f),
\]
where $\partial f$ is oriented as the boundary of the counter-clockwise oriented face $f$.
This number $q_f$ is called the {\em volume weight\/} of the face $f$.

\begin{prop}\label{prop:weights}
The Gibbs measure on $\D(\G)$ given by the edge weight system $w$ translates into the following probability measure on $\mathcal{H}_{D_0}(X,f_0)$:
\[
\mbox{Prob}_{D_0}(h)=\frac{q(h)}{Z_{D_0,f_0}(X,q)},
\]
where
\[
q(h)=\prod_{f\in F(X)}q_f^{h(f)}\quad\text{and}\quad Z_{D_0,f_0}(X;q)=\sum_{h\in \mathcal{H}_{D_0}(X,f_0)}q(h).
\]
Furthermore, this measure is independant of the choice of $f_0$.
Finally, the bijection $\mathcal{H}_{D_0}(X,f_0)\to\mathcal{H}_{D_1}(X,f_0)$ given by $h\mapsto h+h_{D_0,D_1}$ is invariant with respect
to the measures $\mbox{Prob}_{D_0}$ and $\mbox{Prob}_{D_1}$.
\end{prop}
\begin{proof}
For any $D\in\D(\G)$, we have
\begin{align*}
w(D)w(D_0)^{-1}&=\prod_{e\in D}w(e)\prod_{e\in D_0}w(e)^{-1}=w_\beta(C(D,D_0))\\
&=w_\beta(\partial\sigma_{D,D_0})=w_\beta\Big(\sum_{f\in F(X)}h_{D,D_0}(f)\partial f\Big)\\
&=\prod_{f\in F(X)}w_\beta(\partial f)^{h_{D,D_0}(f)}=\prod_{f\in F(X)}q_f^{h_{D,D_0}(f)}=q(h_{D,D_0}).
\end{align*}
The proposition follows easily from this equality.
\end{proof}

Let $V(\G)$ (resp. $E(\G)$) denote the set of vertices (resp. of edges) of $\G$. Recall that the group
\[
{\mathcal G}(\G)=\{s\colon V(\G)\to\R_{>0}\}
\]
acts on the set of weight systems on $\G$ by $(sw)(e)=s(e_+)w(e)s(e_-)$, where $e_+$ and $e_-$ are the two vertices adjacent to
the edge $e$. As observed in Section \ref{sub:dimer-graph}, the Gibbs measure on $\D(\G)$ is invariant under the action of the group
${\mathcal G}(\G)$.

Note also that this action is free unless $\G$ is bipartite. In this later case, the 1-parameter family of elements $s_\lambda\in{\mathcal G}(\G)$
given by $s_\lambda(v)=\lambda$ if $v$ is black and $s_\lambda(v)=\lambda^{-1}$ if $v$ is white
act as the identity on the set of weight systems. Hence, if $\G$ is bipartite, the number of ``essential" parameters is equal to $|E(\G)|-|V(\G)|+1$.
If this bipartite graph is planar, then
\[
|E(\G)|-|V(\G)|+1=|F(X)|-\chi(S^2)+1=|F(X)|-1.
\]
The $|F(X)|$ volume weights $q_f$ are invariant with respect to the action of ${\mathcal G}(\G)$. They can be normalized
in such a way that $\prod_{f\in F(X)}q_f=1$, giving exactly $|F(X)|-1$ parameters. Thus, in the height
function formulation of the Gibbs measure, only essential parameters appear.

\subsection{Height functions for bipartite surface graphs}
\label{sub:height-gen}

Let us now address the general case of a bipartite surface graph $\GSS$, possibly disconnected, and possibly with boundary $\partial\G\subset\partial\SI$.
Fix a family $\gamma=\{\gamma_i\}_{i=1}^{b_1}$ of oriented simple curves in $\G$ representing a basis in $H_1(\SI,\partial\SI;\Z)$.
Note that such a family of curves exists since $\G$ is the 1-squeletton of a cellular decomposition $X$ of $\Sigma$.

Given any $D,D'\in\D(\G,\partial\G)$, the homology class of $C(D,D')=D-D'$ can be written in a unique way
\[
[C(D,D')]=\sum_{i=1}^{b_1}a^\gamma_{D,D'}(i)[\gamma_i]\in H_1(\SI,\partial\SI;\Z),
\]
with $a^\gamma_{D,D'}(i)\in\Z$. Hence, $C(D,D')-\sum_{i=1}^{b_1}a^\gamma_{D,D'}(i)\gamma_i$ is a 1-boundary $(rel\;\partial X)$, that is, there exists
$\sigma^\gamma_{D,D'}\in C_2(X,\partial X;\Z)=C_2(X;\Z)$ such that
\begin{equation}\label{equ:height}
C(D,D')-\partial\sigma^\gamma_{D,D'}-\sum_{i=1}^{b_1}a^\gamma_{D,D'}(i)\gamma_i\;\in\;C_1(\partial X;\Z).
\end{equation}
The 2-cochain $h^\gamma_{D,D'}\in C^2(X;\Z)$ dual to $\sigma^\gamma_{D,D'}$ is called a {\em height function associated to $D,D'$
with respect to $\gamma$\/}. Since $Z_2(X,\partial X;\Z)=H_2(X,\partial X;\Z)=H_2(\SI,\partial\SI;\Z)\cong H^0(\SI;\Z)$, the 2-chain
$\sigma^\gamma_{D,D'}$ is uniquely determined by $D,D'$ and $\gamma$ up to an element of $H^0(\SI;\Z)$, and the same holds for $h^\gamma_{D,D'}$.
In other words, the set of height functions associated to $D,D'$ with respect to $\gamma$ is an affine $H^0(\SI;\Z)$-space:
it admits a freely transitive action of the abelian group $H^0(\SI;\Z)$.
One can normalize the height functions by choosing some family $\mathcal{F}_0$ of faces of $X$, one for each connected component of $X$,
and by setting $h^\gamma_{D,D'}(f_0)=0$ for all $f_0\in\mathcal{F}_0$.

Given $h\in C^2(X;\Z)$, set $C(h)=\partial\sigma\in C_1(X;\Z)$, where $\sigma\in C_2(X;\Z)$ is dual to $h\in C^2(X;\Z)$.
Given a fixed $D_0\in\D(\G,\partial\G)$, let ${\mathcal C}(D_0)$ denote the set of all $C\subset\G$
consisting of disjoint oriented 1-cycles $(rel\;\partial\G)$ such that the following condition holds:
for all $e\in D_0$, either $e$ is contained in $C$ or $e$ is disjoint from $C$.

Finally, let ${\mathcal H}^\gamma_{D_0}(X,\mathcal{F}_0)$ denote
the set of pairs $(h,a)\in C^2(X;\Z)\times\Z^{b_1}$ which satisfy the following properties:
\begin{ticklist}
\item{$h(f_0)=0$ for all $f_0$ in $\mathcal{F}_0$;}
\item{there exists $C\in{\mathcal C}(D_0)$ such that $C-C(h)-\sum_{i=1}^{b_1}a(i)\gamma_i\in C_1(\partial X;\Z)$.}
\end{ticklist}
We obtain the following generalization of Proposition \ref{prop:bij}. The proof is left to the reader.

\begin{prop}
Given any $(h,a)\in\mathcal{H}^\gamma_{D_0}(X,\mathcal{F}_0)$, there is a unique dimer configuration $D\in\D(\G,\partial\G)$ such that
$h^\gamma_{D,D_0}=h$ and $a^\gamma_{D,D_0}=a$.
Furthermore, given any two dimer configuration $D_0,D_1\in\D(\G,\partial\G)$, there is a canonical bijection
\[
\mathcal{H}^\gamma_{D_0}(X,\mathcal{F}_0)\to\mathcal{H}^\gamma_{D_1}(X,\mathcal{F}_0)
\]
given by $(h,a)\mapsto (h+h^\gamma_{D_0,D_1},a+a^\gamma_{D_0,D_1})$.\qed
\end{prop}

Recall that the boundary conditions on dimer configurations induce a partition
\[
\D(\G,\partial\G)=\bigsqcup_{\partial D_0'}\D(\G,\partial\G\,|\,\partial D_0'),
\]
where $\D(\G,\partial\G\,|\,\partial D_0')=\{D\in\D(\G,\partial\G)\,|\,\partial D=\partial D_0'\}$. This partition translates into a partition
of $\mathcal{H}^\gamma_{D_0}(X,\mathcal{F}_0)$ via the bijection $\D(\G,\partial\G)\to\mathcal{H}^\gamma_{D_0}(X,\mathcal{F}_0)$ given by
$D\mapsto(h^\gamma_{D,D_0},a^\gamma_{D,D_0})$. Indeed, let $F_\partial(X)$ denote the set of boundary faces of $X$, that is, the set of faces
of $X$ that are adjacent to $\partial\SI$. The choice of a boundary condition $\partial D_0'$ (together with $\mathcal{F}_0$) determines
$h^\gamma_{D,D_0}(f)$ for all $D$ such that $\partial D=\partial D_0'$ and all $f\in F_\partial(X)$. The actual possible
values of $h^\gamma_{D,D_0}$ on the boundary faces depend on $\gamma$, $D_0$ and $\mathcal{F}_0$; they can be determined explicitely. We shall denote
by $\partial h$ such a value of a height function on boundary faces, and call it a {\em boundary condition} for height functions. In short, we obtain
a partition
\[
\mathcal{H}^\gamma_{D_0}(X,\mathcal{F}_0)=\bigsqcup_{\partial h_0'}\mathcal{H}^\gamma_{D_0}(X,\mathcal{F}_0\,|\,\partial h_0')
\]
indexed by all possible boundary conditions on height functions $h_{D,D_0}^\gamma$. Each boundary condition on dimer configurations
corresponds to one boundary condition on height functions via $D\mapsto h^\gamma_{D,D_0}$.

\medskip

Let us now consider an edge weight system $w$ on the bipartite graph $\G$, and a fixed boundary condition $\partial D'_0$.
Recall that the Gibbs measure for the dimer model on $(\G,\partial\G)$ with weight system $w$ and boundary condition $\partial D'_0$ is given by
\[
\mbox{Prob}(D\,|\,\partial D'_0)=\frac{w(D)}{Z(\G; w\,|\,\partial D'_0)},
\]
where
\[
Z(\G; w\,|\,\partial D'_0)=\sum_{D\in\D(\G,\partial\G\,|\,\partial D'_0)} w(D).
\]

Let us realize $\G$ as a surface graph $\GSS$, fix a dimer configuration $D_0\in\D(\G,\partial\G)$, a family
$\gamma=\{\gamma_i\}$ of oriented simple curves in $\G$ representing a basis in $H_1(\SI,\partial\SI;\Z)$,
and a collection $\mathcal{F}_0$ of faces of the induced cellular decomposition $X$ of $\SI$, one face for each connected component of $X$.
We can use the bijection $\D(\G,\partial\G\,|\,\partial D'_0)\to\mathcal{H}^\gamma_{D_0}(X,\mathcal{F}_0\,|\,\partial h'_0)$
given by $D\mapsto (h^\gamma_{D,D_0},a^\gamma_{D,D_0})$ to translate the Gibbs measure into a probability measure on
$\mathcal{H}^\gamma_{D_0}(X,\mathcal{F}_0\,|\,\partial h'_0)$.

To do so, let us first extend the weight system $w$ to all edges of $X$ by setting $w(e)=1$ for all boundary edges of $X$.
As in the planar case, define $w_\beta\colon C_1(X;\Z)\to\R_{>0}$ as the group homomorphism such that, for any oriented edge $e$ of $X$,
\[
w_\beta(e)=
\begin{cases}
w(e) & \text{if the orientation on $e$ agrees with the bipartite orientation;} \\
w(e)^{-1} & \text{otherwise.}
\end{cases}
\]
Note that this makes sense even for boundary edges where there is no bipartite orientation, as $w(e)=1$ for such edges.
Consider the parameters
\begin{align*}
q_f&=w_\beta(\partial f)\quad\text{for all $f\in F(X)\setminus F_\partial(X)$;}\\
q_i&=w_\beta(\gamma_i)\quad\text{for all $1\le i\le b_1$.}
\end{align*}
We obtain the following generalization of Proposition \ref{prop:weights}:

\begin{prop}\label{prop:weight}
Given an element $(h,a)\in\mathcal{H}^\gamma_{D_0}(X,\mathcal{F}_0)$, set
\[
q(h,a)=\prod_{f\in F(X)\setminus F_\partial(X)}q_f^{h(f)}\;\prod_{1\le i\le b_1}q_i^{a(i)}.
\]
Then, the Gibbs measure for the dimer model on $(\G,\partial\G)$ with weight system $w$ and boundary condition $\partial D'_0$
translates into the following probability measure on $\mathcal{H}^\gamma_{D_0}(X,\mathcal{F}_0\,|\,\partial h'_0)$:
\[
\mbox{Prob}_{D_0}(h,a\,|\,\partial h'_0)=\frac{q(h,a)}{Z^\gamma_{D_0,\mathcal{F}_0}(X;q\,|\,\partial h'_0)},
\]
where
\[
Z^\gamma_{D_0,\mathcal{F}_0}(X;q\,|\,\partial h'_0)=\sum_{(h,a)\in \mathcal{H}^\gamma_{D_0}(X,\mathcal{F}_0\,|\,\partial h'_0)}q(h,a).
\]
Furthermore, the measure is independant of the choice of $\mathcal{F}_0$. Finally,
the bijection $\mathcal{H}^\gamma_{D_0}(X,\mathcal{F}_0\,|\,\partial h'_0)\to\mathcal{H}^\gamma_{D_1}(X,\mathcal{F}_0\,|\,\partial h'_1)$ given by
$(h,a)\mapsto (h+h^\gamma_{D_0,D_1},a+a^\gamma_{D_0,D_1})$ is invariant with respect to the measures $\mbox{Prob}_{D_0}$ and $\mbox{Prob}_{D_1}$.
\end{prop}
\begin{proof}
For any $D\in\D(\G,\partial\G\,|\,\partial D'_0)$, equation (\ref{equ:height}) leads to
\[
w_\beta\big(C(D,D_0)-\partial\sigma^\gamma_{D,D_0}-\textstyle\sum_{i=1}^{b_1}a^\gamma_{D,D_0}(i)\gamma_i\big)=1.
\]
Computing the first term, we get
\[
w_\beta(C(D,D_0))=\prod_{e\in D}w(e)\prod_{e\in D_0}w(e)^{-1}=w(D)w(D_0)^{-1}.
\]
As for the second one,
\[
w_\beta(\partial\sigma^\gamma_{D,D_0})=w_\beta\Big(\sum_{f\in F(X)}h_{D,D_0}(f)\partial f\Big)=\prod_{f\in F(X)}q_f^{h_{D,D_0}(f)}.
\]
Since $w_\beta(\gamma_i)=q_i$, these equations lead to
\[
w(D)=w(D_0)\prod_{f\in F(X)}q_f^{h^\gamma_{D,D_0}(f)}\prod_{1\le i\le b_1}q_i^{a^\gamma_{D,D_0}(i)}=\lambda\cdot q(h^\gamma_{D,D_0},a^\gamma_{D,D_0}),
\]
where $\lambda=w(D_0)\prod_{f\in F_\partial(X)}q_f^{h^\gamma_{D,D_0}(f)}$ depends only on $D_0$ and $\partial D'_0$.
The proposition follows easily from this equality.
\end{proof}

Let us count the number of essential parameters in the dimer model on $(\G,\partial\G)$ with some boundary condition partitioning $\partial\G$
into $(\partial\G)_{nm}\sqcup (\partial\G)_m$, matched and non-matched vertices. We have $|E(\G)|-|(\partial\G)_{nm}|$ edge weights, with an action
of a $(|V(\G)|-|(\partial\G)_{nm}|)$-parameter group. Since $\G$ is bipartite, there is a $b_0(\G)$-parameter subgroup acting as the identity.
Therefore, the number of essential parameters is equal to
\begin{align*}
|E(\G)|-|V(\G)|+b_0(\G)&=|E(X)|-|\partial\G|-|V(X)|+b_0(X)\\
&=|F(X)|-|\partial\G|-\chi(X)+b_0(X)\\
&=|F(X)\setminus F_\partial(X)|+b_1(\SI)-b_2(\SI).
\end{align*}
The numbers $|F(X)\setminus F_\partial(X)|$ and $b_1(\SI)$ correspond to the parameters $q_f$ and $q_i$. Furthermore,
the parameters $q_f$ can be normalized by $\prod_f q_f=1$, the product being on all faces of a given closed component of $\SI$.
Therefore, we obtain exactly the right number of parameters in this height function formulation of the dimer model.

\begin{rem}
Note that all the results of the first part of the present section can be adapted to the general case of a non-necessarily
bipartite surface graph: one simply needs to work with $\Z_2$-coefficients. However, the height function formulation of the dimer model
using volume weights does require a bipartite structure. It is unknown whether a reformulation of the dimer model with the right
number of parameters is possible in the general case.
\end{rem}

\subsection{The dimer quantum field theory on bipartite surface graphs}

Let us now use these results to reformulate the dimer quantum field theory on bipartite graphs. Let $\GSS$ be a bipartite surface graph,
and let $X$ denote the induced cellular decomposition of $\SI$. Fix a dimer configuration $D_0\in\D(\G,\partial\G)$, a family
$\gamma=\{\gamma_i\}$ of oriented simple curves in $\G$ representing a basis in $H_1(\SI,\partial\SI;\Z)$, and a choice $\mathcal{F}_0$
of one face in each connected component of $X$.
\begin{enumerate}
\item{To $\partial X$, assign
\[
Z(\partial X)=\bigotimes_{f\in F_\partial(X)}W,
\]
where $W$ is the complex vector space with basis $\{\alpha_n\}_{n\in\Z}$, and $F_\partial(X)$ denotes the set of faces of $X$ adjacent to the boundary.}
\item{To $X$ with weight system $q=\{q_f\}_{f\in F(X)}\cup\{q_i\}_{1\le i\le b_1(\SI)}$, assign
\[
Z^\gamma_{D_0,\mathcal{F}_0}(X;q)=\sum_{\partial h}Z^\gamma_{D_0,\mathcal{F}_0}(X;q\,|\,\partial h)\,\alpha(\partial h)\,\in\,Z(\partial X),
\]
where
\[
Z^\gamma_{D_0,\mathcal{F}_0}(X;q\,|\,\partial h)=\sum_{(h,a)\in \mathcal{H}^\gamma_{D_0}(X,\mathcal{F}_0\,|\,\partial h)}
\prod_{f\in F(X)\setminus F_\partial(X)}q_f^{h(f)}\;\prod_{1\le i\le b_1(\SI)}q_i^{a(i)}
\]
and $\alpha(\partial h)=\bigotimes_{f\in F_\partial(X)}q_f^{h(f)}\alpha_{h(f)}$.}
\end{enumerate}

Recall the notation $a(\partial D)\in Z(\partial\G)$ of Section \ref{sub:dimerQFT}. The bijection
$\D(\G,\partial\G)\to\mathcal{H}^\gamma_{D_0}(X,\mathcal{F}_0)$ induces an inclusion $j\colon Z(\partial\G)\hookrightarrow Z(\partial X)$ such that
\[
j(a(\partial D))=\bigotimes_{f\in F_\partial(X)}\alpha_{h^\gamma_{D,D_0}(f)}.
\]
Therefore, using the proof of Proposition \ref{prop:weight},
\begin{align*}
j(Z(\G;w))&=\sum_{\partial D}Z(\G;w\,|\,\partial D)\,j(a(\partial D))\\
&=\sum_{\partial h}Z^\gamma_{D_0,\mathcal{F}_0}(X;q\,|\,\partial h)\,w(D_0)
\prod_{f\in F_\partial(X)}q_f^{h(f)}\bigotimes_{f\in F_\partial(X)}\alpha_{h(f)}\\
&=w(D_0)\,Z^\gamma_{D_0,\mathcal{F}_0}(X;q),
\end{align*}
where the weight system $q$ is obtained from $w$ by $q_f=w_\beta(\partial f)$ and $q_i=w_\beta(\gamma_i)$.

In this setting, the gluing axiom makes sense only when the data $\beta$, $D_0$ and $\gamma$ are compatible with the gluing map $\varphi$.
In such a case case, it holds by the equality above and the results of Section \ref{sub:dimerQFT}.

The equivalence between the quantum field theories formulated in Section \ref{sub:FermionsQFT} and in the present section
should be regarded as a discrete version of the boson-fermion correspondence on compact Riemann surfaces (see \cite{AGBMNV}).

\bibliographystyle{amsplain}

\end{document}